\documentclass[aps,prl,reprint,superscriptaddress,longbibliography,floatfix,
nofootinbib,nobibnotes]{revtex4-2}

\usepackage{amsmath,amssymb,mathtools,bm}
\usepackage{amsthm}
\usepackage{booktabs}
\usepackage{microtype}
\usepackage{graphicx}
\usepackage{xcolor}
\usepackage{tikz}
\usetikzlibrary{calc,positioning,fit}
\usepackage[colorlinks=true,linkcolor=blue,citecolor=blue,urlcolor=blue,
bookmarks=false]{hyperref}

\definecolor{electricblue}{RGB}{43,92,151}
\definecolor{magneticred}{RGB}{177,55,48}
\definecolor{criticalgreen}{RGB}{52,120,82}
\definecolor{softgray}{RGB}{95,98,103}
\definecolor{panelblue}{RGB}{235,242,250}
\definecolor{panelred}{RGB}{252,239,237}
\definecolor{panelgreen}{RGB}{237,247,241}
\definecolor{clockorange}{RGB}{205,116,40}
\definecolor{panelorange}{RGB}{253,244,232}
\definecolor{criticalgray}{RGB}{112,115,120}
\definecolor{systemBblue}{RGB}{102,153,204}

\newcommand{\Rep}{\operatorname{Rep}}

\newcommand{\diam}{\operatorname{diam}}

\newcommand{\MBL}{\mathrm{MBL}}

\theoremstyle{definition}

\makeatletter
\newcommand{\hideMainToc}{%
  \let\savedlsection\l@section
  \renewcommand{\l@section}[2]{}%
}
\newcommand{\showMainToc}{\let\l@section\savedlsection}
\makeatother

\hypersetup{
  pdftitle={Emergent Replica Clock Unifies Many-Body Localization and Thermalization},
  pdfauthor={Tong Liu},
  pdfsubject={Replica-clock correlations, the MBL--ETH transition, and quantized replica resolution}
}

\begin{document}

\title{Emergent Replica Clock Unifies Many-Body Localization and Thermalization}

\author{Tong Liu}
\email{t6tong@njupt.edu.cn}
\affiliation{Department of Applied Physics, School of Science,
Nanjing University of Posts and Telecommunications, Nanjing 210003, China}

\date{August 5, 2026}

\begin{abstract}
Many-body localization (MBL) and eigenstate thermalization (ETH) are
traditionally distinguished by a collection of separate diagnostics, not by a
single order parameter.
We show that replication and folding of isolated unitary dynamics generate a
complex statistical mechanics of forward--backward history pairings and, in a
closed cyclic infrared sector, an emergent clock
$r_x\in\mathbb Z_t$. Its correlator has three distinct asymptotics: exponential
decay in MBL, scale-free decay at criticality, and long-range locking in the
thermal phase. A controlled $l$-bit reduction makes the clock action
quasi-local, while its inverse correlation length obeys
$\xi_{\rm pair}^{-1}=\kappa_{\rm el}=\ln|\rho_0/\rho_q|$, unifying history
coherence, the defect line tension, and the transfer spectrum. Replica
order supplies a second, discrete coordinate---the first order at which a
hidden dynamical invariant becomes visible---and can distinguish localized
dynamics that share the same spatial correlation length.
\end{abstract}

\maketitle
\addtocontents{toc}{\protect\setcounter{tocdepth}{-1}}
\addtocontents{toc}{\protect\hideMainToc}
\paragraph{Introduction.---}
The thermalization of an isolated quantum system and its breakdown in
many-body localization (MBL) are usually characterized by complementary
diagnostics: level statistics, entanglement scaling, transport coefficients,
and the persistence of local memory
\cite{Deutsch1991,Srednicki1994,Rigol2008,DAlessio2016,Gornyi2005,
Basko2006,OganesyanHuse2007,PalHuse2010,Bardarson2012,Serbyn2013,Huse2014,
Imbrie2016,NandkishoreHuse2015,Abanin2019,Luitz2015,Vosk2015,Potter2015,
DeRoeck2017,Thiery2018,Khemani2017,Morningstar2022,Szoldra2024,
Schreiber2015,Choi2016MBL,Lukin2019}. Each of these probes captures a
different facet of the transition, but none is a genuine order parameter---a
single statistical variable whose spatial organization directly distinguishes
MBL from the eigenstate thermalization hypothesis (ETH). This absence is
conceptually unsatisfactory and practically limiting: without a unified order
parameter, the nature of the MBL--ETH transition, its critical properties, and
the possible coexistence of distinct localized phases remain obscured.

The missing variable appears when one changes the object of study from
physical states to quantum histories. Observables such as the spectral form
factor, return probability, and multitime amplitudes naturally compare a
unitary history with its complex conjugate, so their mathematical structure is
a pairing between forward and backward histories. Such permutation pairings
are known to organize replicated transfer matrices and spectral correlations
\cite{ChanDeLucaChalker2018PRX,ChanDeLucaChalker2018PRL,
BertiniKosProsen2018,Friedman2019,GarrattChalker2021PRX,
GarrattChalker2021PRL,Lerose2021,Nahum2017,Nahum2018,Keyserlingk2018,
Rakovszky2018,Khemani2018,ZhouNahum2019,Foligno2023}.
Here we promote these pairings from a contraction rule to the configurations
of a statistical mechanics: isolated unitary evolution, after replication and
folding, generates an exact partition sum over history-pairing sectors
\cite{LiChenFisher2019,Skinner2019,Jian2020,Bao2020,HoChoi2022,
ClaeysLamacraft2022,Ippoliti2023,Cotler2023,Choi2023,Mark2024,
MandalClaeysRoy2026,ODonovan2026}.
\begin{figure*}[t]
\centering
\begin{tikzpicture}[
  >=latex,font=\sffamily\scriptsize,
  history/.style={line width=.85pt},
  pair/.style={->,line width=.8pt},
  dot/.style={circle,fill=white,draw=softgray,inner sep=1.25pt}
]
\begin{scope}[xshift=0cm,yshift=3.45cm]
 \filldraw[fill=panelblue,draw=softgray!55,rounded corners=3pt]
   (0,0) rectangle (7.85,3.05);
 \node[anchor=west,font=\bfseries] at (.22,2.78) {(a) Replicate and fold};
 \draw[history,->,electricblue] (.70,.72)--(5.72,.72);
 \draw[history,->,softgray] (5.72,1.78)--(.70,1.78);
 \draw[electricblue,line width=.9pt] (5.72,.72)
   to[out=12,in=-12] (6.12,1.25) to[out=168,in=-168] (5.72,1.78);
 \filldraw[fill=electricblue,draw=white,line width=.5pt] (6.12,1.25) circle (2.5pt);
 \node[electricblue,anchor=north] at (6.12,.97) {fold};
 \node[electricblue,anchor=north] at (3.15,.62) {Forward: $U$};
 \node[softgray,anchor=south] at (3.15,1.88) {Backward: $\overline U$};
 \foreach \x in {1.25,2.18,3.11,4.04,4.97}
   \draw[softgray!65,densely dashed,line width=.55pt] (\x,.87)--(\x,1.63);
 \node[softgray,align=center] at (3.10,2.42)
   {$t$ replicated forward--backward pairs};
\end{scope}
\begin{scope}[xshift=8.25cm,yshift=3.45cm]
 \filldraw[fill=panelred,draw=softgray!55,rounded corners=3pt]
   (0,0) rectangle (7.85,3.05);
 \node[anchor=west,font=\bfseries] at (.22,2.78) {(b) Full permutation sector $S_t$};
 \foreach \x/\lab in {.75/1,2.05/2,3.35/3,4.65/4,5.95/t}
   {\node[dot] (fb\lab) at (\x,.55) {};
    \node[dot] (bb\lab) at (\x,2.18) {};
    \node[below=1pt] at (\x,.43) {$F_{\lab}$};
    \node[above=1pt] at (\x,2.30) {$B_{\lab}$};}
 \draw[pair,electricblue] (fb1)--(bb4);
 \draw[pair,magneticred] (fb2)--(bbt);
 \draw[pair,softgray] (fb3)--(bb1);
 \draw[pair,criticalgreen] (fb4)--(bb2);
 \draw[pair,magneticred!70!softgray] (fbt)--(bb3);
 \node[fill=white,rounded corners=1pt,inner sep=1.5pt] at (3.35,1.40)
   {noncyclic pairings};
 \draw[magneticred,line width=2.0pt] (6.65,.86)--(7.35,1.56);
 \draw[magneticred,line width=2.0pt] (6.65,1.56)--(7.35,.86);
 \node[magneticred,align=center] at (7.00,2.15) {IR\\suppressed};
\end{scope}
\begin{scope}[xshift=0cm,yshift=0cm]
 \filldraw[fill=panelgreen,draw=softgray!55,rounded corners=3pt]
   (0,0) rectangle (7.85,3.05);
 \node[anchor=west,font=\bfseries] at (.22,2.78) {(c) Cyclic infrared branch};
 \foreach \x/\lab in {.85/1,2.15/2,3.45/3,4.75/4,6.05/t}
   {\node[dot] (fc\lab) at (\x,.55) {};
    \node[dot] (bc\lab) at (\x,2.12) {};
    \node[below=1pt] at (\x,.43) {$F_{\lab}$};
    \node[above=1pt] at (\x,2.24) {$B_{\lab}$};}
 \draw[pair,electricblue] (fc1)--(bc2);
 \draw[pair,criticalgreen] (fc2)--(bc3);
 \draw[pair,electricblue] (fc3)--(bc4);
 \draw[pair,criticalgreen] (fc4)--(bct);
 \draw[pair,electricblue,rounded corners=4pt]
   (fct)--(6.78,.55)--(6.78,.14)--(.45,.14)--(.45,2.12)--(bc1);
 \node[fill=white,rounded corners=1pt,inner sep=1.5pt] at (3.48,1.33)
   {$\sigma_x=c^{r_x}$};
\end{scope}
\begin{scope}[xshift=8.25cm,yshift=0cm]
 \filldraw[fill=panelorange,draw=softgray!55,rounded corners=3pt]
   (0,0) rectangle (7.85,3.05);
 \node[anchor=west,font=\bfseries] at (.22,2.78) {(d) Emergent replica clock};
 \coordinate (O) at (2.15,1.35);
 \draw[clockorange,line width=1.1pt,fill=white] (O) circle (.86);
 \foreach \a in {0,45,...,315}
   {\draw[clockorange,line width=.65pt]
      ($(O)+({.70*cos(\a)},{.70*sin(\a)})$)--
      ($(O)+({.84*cos(\a)},{.84*sin(\a)})$);}
 \fill[clockorange] (O) circle (1.4pt);
 \draw[->,clockorange,line width=1.2pt] (O)--++(48:.64);
 \node[clockorange,font=\bfseries] at (2.15,.20) {$r_x\in\mathbb Z_t$};
 \node[align=left,anchor=west] at (3.55,1.64)
   {relative history frame};
 \node[align=left,anchor=west,font=\normalsize] at (3.55,1.04)
   {$F_a\longleftrightarrow B_{a+r_x\ ({\rm mod}\ t)}$};
\end{scope}
\end{tikzpicture}
\caption{\textbf{Birth of the replica clock.}
(a) Replication and folding place forward and backward unitary histories in
one local tensor network. (b) The resulting $S_t$ pairing sector contains
noncyclic crossings, which are gapped or dephase under infrared coarse
graining. (c) The surviving coherent pairings form the cyclic branch
$\sigma_x=c^{r_x}$. (d) Its local coordinate is the replica clock
$r_x\in\mathbb Z_t$.}
\label{fig:architecture}
\end{figure*}
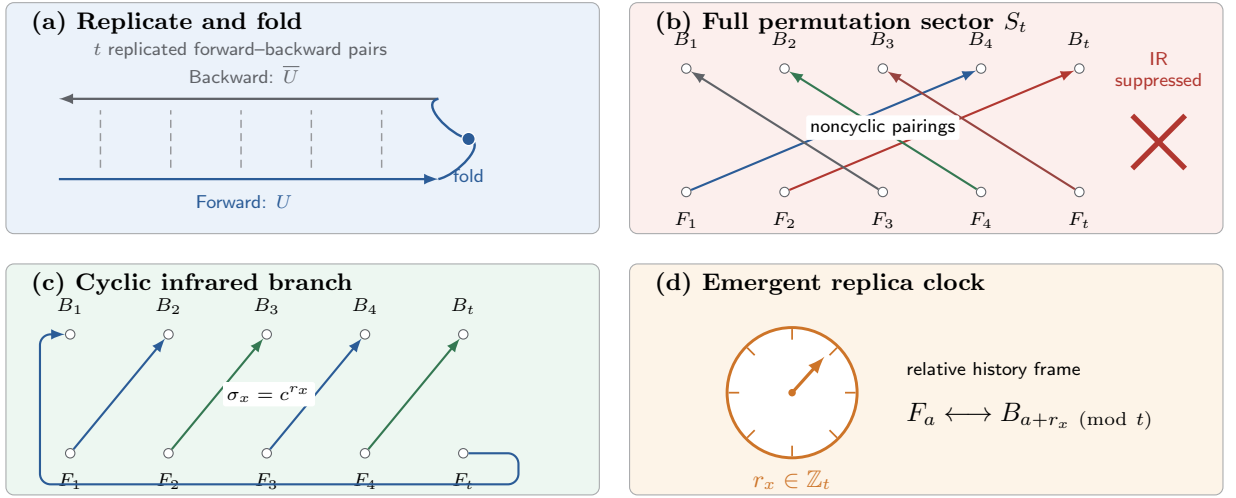

In the infrared limit, the coherent branch of this history-pairing ensemble is
a cyclic clock field $r_x\in\mathbb Z_t$. Its two-point correlator $C_t(R)$
exhibits exactly three asymptotic regimes: exponential decay (MBL), power-law
decay (critical), and long-range locking (ETH). An $l$-bit reduction welds this
clock geometry to microscopic dynamics, yielding a quasi-local effective
action whose transfer matrix gives
$\xi_{\rm pair}^{-1}=\kappa_{\rm el}=\ln|\rho_0/\rho_q|$---a single identity
that unifies history correlation length, defect line tension, and the transfer
spectrum. Beyond spatial phases, replica order $t$ provides a discrete
resolution of many-body memory, distinguishing localized dynamics with
identical $\xi_{\rm pair}$. The construction is model independent and applies
to any local unitary dynamics with a closed cyclic infrared sector.

\paragraph{From isolated dynamics to a replica clock.---}
Vectorizing the channel $\mathcal E_U(X)=UXU^\dagger$ converts one
forward--backward pair into $U\otimes\overline U$. At replica order $t$, the
folded channel is, up to reshuffling,
$U^{\otimes t}\otimes\overline U^{\,\otimes t}$. Locality factorizes it into
doubled-replica tensors. Whenever a pairing sector $\Omega_t\subseteq S_t$ is
closed by those tensors, a Gram-dual resolution on internal cuts gives the
exact configuration sum
\begin{equation}
 \mathcal G_t=\sum_{\sigma\in\Omega_t}
 \exp\!\left[-S_t^{\mathrm R}[\sigma]+i\Phi_t[\sigma]\right].
 \label{eq:history-sum}
\end{equation}
Thus $\Omega_t$ is the configuration space, $S_t^{\mathrm R}$ its real
statistical cost, and $\Phi_t$ the coherent phase left by the same isolated
dynamics; no thermal reservoir is introduced. The construction and its
closure criterion are derived in Supplemental Material (SM), Secs.~SI and
SII~\cite{SM}.
The configuration-resolved theory is
\begin{equation}
 \mathfrak R_t[U]=(\mathfrak K_t,\mathfrak C_t[U],\mathfrak P_t[U]),
 \qquad \mathfrak R[U]=\{\mathfrak R_t[U]\}_{t\ge1},
 \label{eq:fixed-and-family}
\end{equation}
separating pairing kinematics, constitutive response, and invariant replica
phase before any final contraction.

The field $\sigma_x\in\Omega_t\subseteq S_t$ in Eq.~\eqref{eq:history-sum} is,
in principle, an arbitrary admissible permutation. A classification of
thermodynamic phases, however, is controlled by the infrared (long-wavelength)
limit: only there do finite correlation length, scale invariance, and
long-range order become sharply distinct asymptotic regimes. We therefore
coarse-grain the history-pairing configurations over scales much larger than
the microscopic lattice spacing. Under the cyclic closure that follows from this infrared reduction, all noncyclic pairing sectors---those not belonging to the selected
$C_t$ subgroup---acquire either real weights that are gapped in the transfer
spectrum or rapidly oscillating phases that cancel under coarse graining. The
only branch that remains both coherent and closed under the folded dynamics is
the Abelian cyclic subgroup $C_t\subset S_t$. Writing $c=(1\,2\cdots t)$, the
surviving local pairing is
\begin{equation}
 \sigma_x=c^{r_x},\qquad r_x\in\mathbb Z_t,\qquad
 F_a\longleftrightarrow B_{a+r_x\ (\mathrm{mod}\ t)} .
 \label{eq:clock-field}
\end{equation}
The integer $r_x$ is the \emph{replica-pairing clock}: it records the relative
origin of forward and backward histories, not physical time. Figure
\ref{fig:architecture} shows its microscopic origin and the infrared reduction
that selects the cyclic branch.

\begin{figure}[t]
\centering
\begin{tikzpicture}[>=latex,font=\sffamily\scriptsize]
 \filldraw[fill=white,draw=softgray!55,rounded corners=2pt]
   (.15,3.05) rectangle (7.35,6.25);
 \node[anchor=west,font=\bfseries] at (.30,5.98) {(a) Clock correlations};
 \draw[->,softgray,line width=.65pt] (1.15,3.38)--(7.15,3.38)
   node[below=2pt] {$R$};
 \draw[->,softgray,line width=.65pt] (1.15,3.38)--(1.15,5.78);
 \node[rotate=90,softgray] at (.48,4.58) {$|C_{1,t}(R)|$ (log scale)};
 \draw[electricblue,very thick] plot[smooth] coordinates
   {(1.18,5.55)(1.75,4.98)(2.35,4.40)(2.95,3.92)(3.55,3.63)(4.20,3.52)(4.80,3.47)(5.70,3.43)(6.80,3.40)};
 \draw[criticalgray,very thick,dashed] plot[smooth] coordinates
   {(1.18,5.55)(1.75,5.15)(2.40,4.85)(3.20,4.56)(4.20,4.29)(5.45,4.05)(7.05,3.84)};
 \draw[magneticred,very thick,dash pattern=on 5pt off 1.7pt on .8pt off 1.7pt] plot[smooth] coordinates
   {(1.18,5.55)(1.75,5.13)(2.40,4.87)(3.20,4.73)(4.20,4.68)(5.45,4.67)(7.05,4.67)};
 \draw[<->,electricblue,line width=.65pt] (1.25,3.61)--(3.15,3.61);
 \node[electricblue,fill=white,inner sep=1pt] at (2.20,3.61) {$\xi_{\rm pair}$};
 \draw[<->,magneticred,line width=.65pt] (6.82,3.43)--(6.82,4.64);
 \node[magneticred,fill=white,inner sep=1pt,anchor=east] at (6.76,4.10) {$m_t^2$};
 \node[electricblue,anchor=west] at (3.28,3.92) {MBL: $e^{-R/\xi_{\rm pair}}$};
 \node[criticalgray,anchor=west] at (4.82,4.32) {critical: $R^{-\eta}$};
 \node[magneticred,anchor=west] at (5.10,4.86) {ETH: $m_t^2>0$};
 \draw[softgray!75,densely dashed,rounded corners=2pt]
   (.15,.10) rectangle (7.35,2.82);
 \node[anchor=west,font=\bfseries] at (.30,2.57) {(b) Correlator--response matrix};
 \fill[panelblue] (1.60,1.94) rectangle (3.42,2.34);
 \fill[softgray!10] (3.42,1.94) rectangle (5.24,2.34);
 \fill[panelred] (5.24,1.94) rectangle (7.06,2.34);
 \node[electricblue,font=\bfseries] at (2.51,2.14) {MBL};
 \node[criticalgray,font=\bfseries] at (4.33,2.14) {Critical};
 \node[magneticred,font=\bfseries] at (6.15,2.14) {ETH};
 \fill[electricblue!7] (1.60,1.50) rectangle (3.42,1.94);
 \fill[softgray!7] (3.42,1.50) rectangle (5.24,1.94);
 \fill[magneticred!7] (5.24,1.50) rectangle (7.06,1.94);
 \node[softgray,anchor=east] at (1.48,1.72) {$C_t(R)$};
 \node[electricblue] at (2.51,1.72) {$e^{-R/\xi_{\rm pair}}$};
 \node[criticalgray] at (4.33,1.72) {$R^{-\eta}$};
 \node[magneticred] at (6.15,1.72) {$m_t^2>0$};
 \fill[electricblue!86!black] (1.60,.84) rectangle (3.42,1.50);
 \fill[softgray!8] (3.42,.84) rectangle (5.24,1.50);
 \fill[softgray!8] (5.24,.84) rectangle (7.06,1.50);
 \node[softgray,anchor=east] at (1.48,1.17) {$\kappa_{\rm el}$};
 \node[white,font=\bfseries] at (2.51,1.17) {$>0$};
 \node[criticalgray,font=\bfseries] at (4.33,1.17) {$0$};
 \node[criticalgray,font=\bfseries] at (6.15,1.17) {$0$};
 \fill[softgray!8] (1.60,.18) rectangle (3.42,.84);
 \fill[softgray!8] (3.42,.18) rectangle (5.24,.84);
 \fill[magneticred!86!black] (5.24,.18) rectangle (7.06,.84);
 \node[softgray,anchor=east] at (1.48,.51) {$\kappa_{\rm twist}$};
 \node[criticalgray,font=\bfseries] at (2.51,.51) {$0$};
 \node[criticalgray,font=\bfseries] at (4.33,.51) {$0$};
 \node[white,font=\bfseries] at (6.15,.51) {$>0$};
 \foreach \x in {1.60,3.42,5.24,7.06}
   \draw[softgray!38,line width=.35pt] (\x,.18)--(\x,2.34);
 \foreach \y in {.18,.84,1.50,1.94,2.34}
   \draw[softgray!38,line width=.35pt] (1.60,\y)--(7.06,\y);
\end{tikzpicture}
\caption{\textbf{Three clock regimes and their two responses.}
(a) Color and line style distinguish exponential decay (MBL), scale-free
decay (critical), and long-range locking (ETH). (b) The response matrix
directly pairs each correlator with its two asymptotic costs: MBL has line
tension only, ETH has twist stiffness only, and both vanish at criticality.}
\label{fig:clock-phases}
\end{figure}
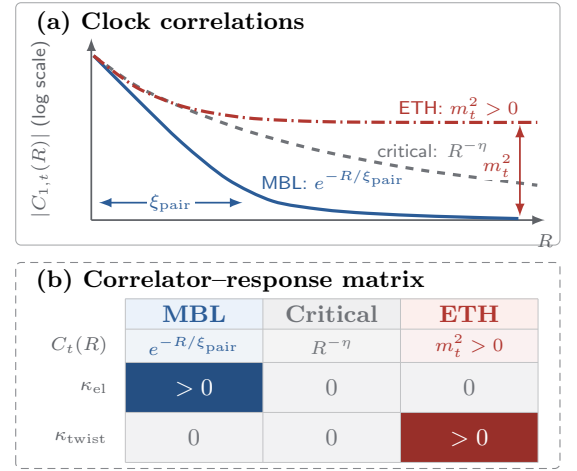

\paragraph{One correlator, three regimes.---}
For a nontrivial character $q\in\widehat{\mathbb Z}_t$, define
\begin{equation}
 C_{q,t}(R)=\left\langle
 \exp\!\left[\frac{2\pi iq}{t}(r_x-r_{x+R})\right]
 \right\rangle .
 \label{eq:clock-correlator}
\end{equation}
This is a correlator of relative history frames, not of microscopic spins.
Let $\xi_{\rm pair}$ be the exponential correlation length of $C_{1,t}$ and
$m_t^2=\lim_{R\to\infty}|C_{1,t}(R)|$. Within a faithful cyclic sector, the
history-space classification is
\begin{equation}
 \MBL_{\rm hist}\Longleftrightarrow
 0<\xi_{\rm pair}<\infty,\qquad m_t=0,
 \label{eq:main-new-mbl-definition}
\end{equation}
with the decoupled $\xi_{\rm pair}=0$ point included as the ideal localized
limit. Criticality has $\xi_{\rm pair}=\infty$ but $m_t=0$; ETH has $m_t>0$.
The counterintuitive clock order in ETH is order in \emph{pairing space}:
scrambling erases independent local memories, leaving the dominant
long-distance contractions locked to one cyclic origin. Figure
\ref{fig:clock-phases} combines this geometrical classification with its two
conjugate defect responses.

The two response coefficients in Fig.~\ref{fig:clock-phases} have a direct
physical interpretation: together they quantify the cost of twisting the
clock frame locally between two endpoints and globally across a winding seam.
In MBL, $\kappa_{\rm el}>0$ means that maintaining a
coherent relative history frame between distant endpoints costs a free energy
proportional to their separation: history alignment is expensive. Yet
$\kappa_{\rm twist}=0$ means that a system-spanning boundary twist can be
absorbed without an extensive free-energy cost. Independent local memories
therefore make the clock medium soft against global rigidity. In ETH the
response is inverted. Scrambling erases independent local memory, so distant
histories align without a line tension, $\kappa_{\rm el}=0$; but long-range
locking forces the clocks into one common frame, making a winding boundary
twist costly, $\kappa_{\rm twist}>0$. The transition thus exchanges a line
tension for a twist stiffness.

\paragraph{Physical anchor and the three-language identity.---}
The two response coefficients in Fig.~\ref{fig:clock-phases} are obtained
from the transfer matrix of Eq.~\eqref{eq:lbit-clock-action}:
$\kappa_{\rm el}$ from open-endpoint propagation and $\kappa_{\rm twist}$ from
twisted boundary conditions.
The classification becomes physical only if the retained clock is tied to
the microscopic localized dynamics. In a controlled $l$-bit regime,
\begin{equation}
 H_{\rm lbit}=\sum_X J_X\tau_X^z,
 \qquad |J_X|\le J_0e^{-\diam(X)/\xi_{\rm lbit}},
 \label{eq:lbit-hamiltonian}
\end{equation}
folding and restricting to the cyclic branch give a quasi-local clock action
\begin{equation}
 \begin{aligned}
 S_{\rm eff}^{(t)}[r]
 &=\sum_{i<j}K_{ij}^{(t)}(1-\delta_{r_i,r_j})
   +\sum_{|X|>2}K_X^{(t)}[r_X],\\[-1mm]
 |K_X^{(t)}|&\le A_t e^{-\diam(X)/\xi_{\rm lbit}} .
\end{aligned}
\label{eq:lbit-clock-action}
\end{equation}
The coefficients are folded-history response functions, not the bare signed
$J_X$. Its leading two-body sector is a spatially decaying random-bond
$\mathbb Z_t$ Potts/clock model,
$K_{ij}^{(t)}\sim e^{-|i-j|/\xi_{\rm lbit}}$; the higher-body terms preserve
the same cyclic symmetry and quasi-locality. In this language, the clock order
parameter distinguishes a paramagnetic (MBL) from a ferromagnetic (ETH)
ordering of the history-pairing field. In the localized
cluster-expansion regime, the inherited inhomogeneity places this history
clock in a disordered, paramagnetic or glassy sector whose two-point function
decays exponentially. The resulting $\xi_{\rm pair}$ is therefore an output:
the renormalized correlation length of the clock model. When a coherent
ferromagnetic response becomes long ranged, the same clock may undergo a
continuous ordering transition, realizing ETH as long-range order in history
space. Growth of the bare $l$-bit range promotes this instability but is not,
by itself, sufficient to guarantee it.

Diagonalizing the spatial transfer operator gives the leading eigenvalues
$\rho_0^{(t)}$ in the trivial sector and $\rho_q^{(t)}$ in clock-character
sector $q$; their ratio fixes the asymptotic decay. Hence, when both modes are
isolated and have nonzero source overlap,
\begin{equation}
 \xi_{q,t}^{-1}=\kappa_{{\rm el},q}^{(t)}
 =\ln\frac{|\rho_0^{(t)}|}{|\rho_q^{(t)}|}.
 \label{eq:tension-transfer}
\end{equation}
\textbf{This is the central identity of the work:} the clock correlation length, the line tension
for twisting a relative history frame between two endpoints, and the
transfer-matrix gap are not independent---they are the same physical scale.
For disordered transfer products the last term is the corresponding
difference of Lyapunov exponents. Thus $\xi_{\rm pair}$ is not guessed: it is
the correlation length extracted from the transfer spectrum of the $l$-bit
clock action. The derivation, including why no generic numerical equality with
the bare decay length of every $J_X$ is required, is given in SM,
Sec.~SIV D.

Imposing a relative shift across a system-spanning seam,
$r_{x+L}=r_x+m$, defines a conjugate twist stiffness
$\kappa_{\rm twist}$, obtained from the same transfer matrix with twisted
boundary conditions (see SM, Sec.~SIV).

\paragraph{Quantized replica resolution.---}
The discussion above focuses on the magnitude of $\mathcal G_t$, which
determines $\xi_{\rm pair}$. The phase $\Phi_t[\sigma]$ in
Eq.~\eqref{eq:history-sum} carries complementary information, including
topological winding numbers and interference phases between pairing sectors.
Although these phases do not change the leading three-regime classification,
they supply additional invariants for the replica hierarchy. In a cyclic
first-visible sector, the conjugate clock and Fourier entropies obey
$H_{\rm clock}+H_{\rm Fourier}\ge\ln t_\star$ (see SM,
Sec.~SIII F).

Replicated histories carry information beyond the spatial correlation length.
Let $\mathcal A_t$ denote the dynamical invariants
accessible at order $t$. For an information class $\mathcal I$, define
\begin{equation}
 t_\star(\mathcal I)=
 \min\{t:\mathcal I\in\Rep(\mathcal A_t)\}.
 \label{eq:tstar}
\end{equation}
The origin of this discrete resolution is the permutation structure of
replicated histories. At order $t$, the accessible invariant algebra
$\mathcal A_t$ contains only observables compatible with the symmetry of the
$t$-cycle pairings. A dynamical invariant $\mathcal I$---for example, a
specific many-body correlator or a winding number---may therefore require a
minimal order $t_\star$ before it can be represented. This is not a matter of
measurement precision: below $t_\star$, the invariant simply does not exist
in the algebra of admissible observables.

\begin{figure}[!t]
\centering
\begin{tikzpicture}[>=latex,font=\sffamily\scriptsize]
 \filldraw[fill=panelblue,draw=softgray!55,rounded corners=2pt]
   (.12,3.18) rectangle (7.32,5.72);
 \node[anchor=west,font=\bfseries] at (.28,5.46) {(a) Low order: $t=2$};
 \draw[->,softgray,line width=.6pt] (.92,3.55)--(7.08,3.55);
 \draw[->,softgray,line width=.6pt] (.92,3.55)--(.92,5.24);
 \node[rotate=90,softgray] at (.43,4.38) {$|C_{1,t}(R)|$};
 \draw[electricblue,very thick] plot[smooth] coordinates
   {(.98,5.13)(1.65,4.85)(2.35,4.56)(3.10,4.29)(3.90,4.05)(4.80,3.84)(5.80,3.69)(6.92,3.59)};
 \node[electricblue,fill=white,inner sep=1.5pt] at (4.18,4.47)
   {$A=B$};
 \node[softgray,anchor=east] at (6.88,5.16) {indistinguishable};
 \filldraw[fill=white,draw=softgray!55,rounded corners=2pt]
   (.12,.15) rectangle (7.32,2.88);
 \node[anchor=west,font=\bfseries] at (.28,2.62) {(b) First-visible order: $t=t_\star$};
 \draw[->,softgray,line width=.6pt] (.92,.51)--(7.08,.51) node[below=2pt] {$R$};
 \draw[->,softgray,line width=.6pt] (.92,.51)--(.92,2.37);
 \node[rotate=90,softgray] at (.43,1.43) {$|C_{1,t}(R)|$};
 \draw[electricblue,very thick] plot[smooth] coordinates
   {(.98,2.24)(1.65,1.96)(2.35,1.67)(3.10,1.40)(3.90,1.16)(4.80,.95)(5.80,.80)(6.92,.70)};
 \draw[systemBblue,very thick,densely dashed] plot[smooth] coordinates
   {(.98,2.24)(1.65,1.96)(2.35,1.67)(3.10,1.40)(3.70,1.27)(4.25,1.36)(4.80,1.42)(5.40,1.33)(6.10,1.19)(6.92,1.06)};
 \foreach \p in {(4.25,1.36),(4.80,1.42),(5.40,1.33),(6.10,1.19)}
   \filldraw[fill=white,draw=systemBblue,line width=.75pt] \p circle (1.45pt);
 \node[electricblue,anchor=west] at (5.72,.66) {$A$};
 \node[systemBblue,anchor=west] at (6.18,1.28) {$B$};
 \node[systemBblue,align=center] at (4.18,2.08) {first split};
 \draw[->,systemBblue,line width=.75pt] (4.18,1.90)--(4.33,1.48);
 \node[softgray,anchor=west] at (1.18,.72) {same $\xi_{\rm pair}$};
\end{tikzpicture}
\caption{\textbf{Quantized resolution.}
(a) At low replica order, localized dynamics $A$ and $B$ have the same
correlator and are operationally indistinguishable. (b) At $t=t_\star$, a
newly representable invariant produces their first higher-order split even
though the spatial decay scale $\xi_{\rm pair}$ remains the same.}
\label{fig:resolution}
\end{figure}
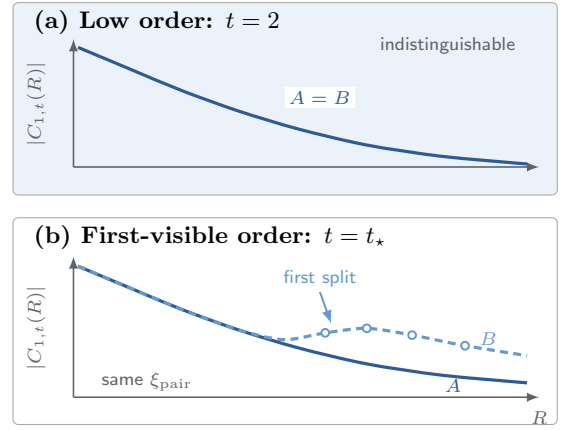

Figure~\ref{fig:resolution} illustrates the operational consequence. Consider
two MBL dynamics $A$ and $B$ that share the same spatial correlation length
$\xi_{\rm pair}$ and hence belong to the same thermodynamic phase. At low
replica order $t<t_\star$, their invariant fingerprints
$\mathcal F_t[A]$ and $\mathcal F_t[B]$ are identical: all accessible
observables agree. At $t=t_\star$, however, an invariant representable only
at this order first appears. If its evaluation differs between the two
dynamics, the curves split---not because their spatial decay has changed, but
because their higher-order history correlations differ. The split is sharp:
it occurs at a discrete threshold rather than through a gradual crossover.

These resolution shells are not thermodynamic phases and require no
free-energy singularity. Instead they refine localized dynamics by adding a
second coordinate $(\xi_{\rm pair},t_\star)$ to the phase diagram. Systems
with identical spatial correlations can thus differ in their many-body
correlation depth, as occurs naturally in $l$-bit models whose multibody
couplings first activate invariants at higher replica orders. The exact
visibility result and its cyclic uncertainty relation are proved in SM,
Sec.~SIII.

\paragraph{Conclusion.---}
This framework reduces the MBL--ETH transition to a geometric statement about
a single object: how far the replica clock remains coherent. The $l$-bit
reduction anchors this geometry to microscopic dynamics, and the transfer
identity $\xi_{\rm pair}^{-1}=\kappa_{\rm el}=\ln|\rho_0/\rho_q|$ exposes its
threefold physical face. Beyond spatial phases, the discrete replica order
$t$ provides a quantized resolution of hidden many-body memory, distinguishing
localized dynamics that are otherwise identical in all low-order diagnostics.
The conjugate defect sector, including its twisted-boundary response, is
detailed in SM, Sec.~SIV. Because the construction uses only locality and a
closed cyclic infrared sector, it is independent of the detailed form of $U$
and applies broadly to unitary dynamics with local structure. The discrete
replica resolution $t_\star$ reveals that quantum information can be hidden in
higher-order history correlations, accessible only at sufficient algebraic
depth. This suggests a
unifying perspective on information retention in quantum many-body systems,
from disordered spin chains to dual-unitary circuits and, potentially, to
holographic models of black-hole evaporation.

\end{document}